\documentclass[twocolumn,showpacs]{revtex4}
\usepackage{amsmath,amsfonts,amssymb}
\usepackage{amsmath}
\usepackage{dcolumn}
\usepackage[pdftex]{graphicx}
\usepackage{bbm}
\usepackage{bm}
\PassOptionsToPackage{caption=false}{subfig}
\usepackage{subfig} % Figures broken into subfigures

\newtheorem{theorem}{Theorem}
\newtheorem{definition}[theorem]{Definition}

\begin{document}

\title{Exact renormalization in quantum spin chains}
\author{Hong-Hao Tu and Mikel Sanz}
\affiliation{Max-Planck-Institut f\"ur Quantenoptik, Hans-Kopfermann-Str. 1, 85748
Garching, Germany}
\date{\today}

\begin{abstract}
We introduce a real-space exact renormalization group method to find exactly
solvable quantum spin chains and their ground states. This method allows us
to provide a complete list for exact solutions within SU(2) symmetric
quantum spin chains with $S\leq 4$ and nearest-neighbor interactions, as
well as examples with $S=5$. We obtain two classes of solutions: One of them
converges to the fixed points of renormalization group and the ground states
are matrix product states. Another one does not have renormalization fixed
points and the ground states are partially ferromagnetic states.
\end{abstract}

\maketitle

\section{Introduction}

Understanding the physical properties of quantum many-body systems is an
important common issue in condensed matter physics and quantum information
theory. The number of parameters required to describe a random state grows
exponentially with the number of particles, which makes the computation of
many-body systems very difficult, even numerically \cite{Cirac-review}.
However, recent development in quantum information theory implies that only
a corner of such a huge Hilbert space is relevant for describing the
low-energy states of physical systems \cite{Verstraete-2006,Hastings-2007}.
The characteristic feature of this corner seems to be an area law \cite%
{Eisert-2010}: the von Neumann entropy of a subsystem in the many-body
ground state scales with the border area, rather than the volume -- the case
for a random state. This means that the ground states of quantum many-body
systems usually only contain a small amount of entanglement. It is natural
to take this advantage and design clever parametrizations of states which
both capture the essential physics and allow classical simulations with a
polynomial time. In one dimension (1D), the matrix product state (MPS) \cite%
{Fannes-1989,Klumper-1991,Cirac-2007}\ is a candidate for such a purpose.
The MPS lies at the heart of the success of the density-matrix
renormalization group (DMRG) \cite%
{White-DMRG,Schollwock-review,Ostlund-1995,Vidal-2004,Verstraet-Cirac-DMRG},
which\ has been proved to be an accurate numerical method for describing the
low-energy states of quantum lattice models. Recently, there are many
interesting extensions along this direction, including infinite MPS\ for
critical systems \cite{Cirac-Sierra-2010}, continuous MPS for quantum field
theories \cite{Cirac-Verstraete-2010}, and projected entangled pair state
(PEPS) for higher dimensional systems \cite%
{Cirac-Verstraete-2004,Sierra-1998,Nishino-2001}.

The MPS also appears to be the exact ground states of certain spin models.
For example, the valence-bond solid (VBS) ground states\ of the
Affleck-Kennedy-Lieb-Tasaki (AKLT) models \cite{Affleck-1987} are matrix
product states. They provide a clear physical picture to the Haldane gap
phenomena \cite{Haldane-1983} and shed light on their \textquotedblleft
nearby" integer-spin Heisenberg antiferromagnets \cite{Arovas-1988}. In
condensed matter physics, the Hamiltonians usually arise with two-body
interactions and SU(2) symmetry since they are relevant to describe
realistic materials. In order to study such systems, a method has been
suggested in Ref. \cite{Sanz-2009} to construct the SU(2) symmetric two-body
parent Hamiltonians for MPS. However, when starting with the Hamiltonians,
in principle it can be extremely hard to find their matrix product ground
states \cite{Norbert-2008}.

The purpose of this paper is to investigate a real-space renormalization
group and its applications in a systematical search for exactly solvable
quantum spin chains. The present approach complements the parent Hamiltonian
method in Ref. \cite{Sanz-2009}, such that one can start from the
Hamiltonians and search for exactly solvable ones. We first briefly review
the basics of real-space renormalization and its extension to systems with
SU(2) symmetry. The presence of symmetry allows us to design a simple exact
renormalization scheme. By using this method, we study quantum spin chains
with SU(2) symmetry and nearest-neighbor interactions. For $S\leq 4$, we
provide complete solutions for the models which are frustration-free for two
neighboring spins. Moreover, we also provide a new MPS solution of $S=5$
which was not previously known. We discuss these exact solutions by dividing
them into two different classes, whose ground states are matrix product
states and partially ferromagnetic states, respectively.

\section{Real-space exact\ renormalization}

Let us consider a chain with $N$ local $d$-dimensional Hilbert spaces $%
\mathcal{H}$, that we can assume local spins. We denote by $|M\rangle \in $ $%
\mathcal{H}$ an orthonormal basis in $\mathcal{H}$. And let us also consider
a translationally invariant Hamiltonian $H=\sum_{i}h_{i}^{(k)}$ containing
local interaction terms acting on contiguous $k$ sites. We can assume
positive semidefinite interaction $h_{i}\geq 0$, since they can always be
achieved by shifting local energy level.

Let us now briefly explain the real-space renormalization process. We start
by coupling the first two spins, whose Hilbert space $\mathcal{H}\otimes
\mathcal{H}\rightarrow \mathcal{H}_{2}$ is mapped into a Hilbert space $%
\mathcal{H}_{2}$ which has in general a dimension $D_{2}<d^{2}$. The
criteria followed to perform this reduction is to conserve only the low
energy states of the Hamiltonian. In general, the method works by finding
the mappings $\mathcal{A}^{[i]}:\mathcal{H}_{i-1}\otimes \mathcal{H}%
\rightarrow \mathcal{H}_{i}$ which carry out this process. We continue this
renormalization procedure until reaching the end of the chain and getting an
orthonormal basis $\{|\chi \rangle \}_{\chi =1}^{D_{N}}$ of the Hilbert
space $\mathcal{H}_{N}$.

Let us show the real-space renormalization process from the $(i-1)$-th spin
to the $i$-th spin, which can be written in a basis as \cite{Ostlund-1995}%
\begin{equation}
|\beta \lbrack i]\rangle =\sum_{\alpha ,M}A_{\alpha ,\beta }^{[M]}|\alpha
\lbrack i-1]\rangle \otimes |M[i]\rangle  \label{eq:isometry}
\end{equation}%
where the input state $|\alpha \lbrack i-1]\rangle \in \mathcal{H}_{i-1}$,
the output state $|\beta \lbrack i]\rangle \in \mathcal{H}_{i}$, and the
Kraus operator $A^{[M_{i}]}$'s are $D_{i-1}\times D_{i}$ matrices satisfying
isometry condition $\sum_{M}A^{[M]\dagger }A^{[M]}=\mathbbm{1}$. Here we
define $D_{0}=1$ so that the Kraus operator $A^{[M_{1}]}$ for the first spin
can be viewed as a row vector.

Equation (\ref{eq:isometry}) shows the real-space renormalization results in an
orthonormal basis $|\chi \rangle $ ($\chi =1\cdots D_{N}$) with a matrix
product form (See Fig. 1a)%
\begin{equation}
|\chi \rangle =\sum_{M_{1}\ldots M_{N}}(A^{[M_{1}]}A^{[M_{2}]}\cdots
A^{[M_{N}]})_{\chi }|M_{1},M_{2}\cdots M_{N}\rangle  \label{eq:MPS}
\end{equation}%
where $D=\max_{i}D_{i}$ is called the \textit{bond dimension} of the matrix
product. In the DMRG algorithm, these matrix product states are used
variationally to find the best approximation of the low energy sector of the
1D systems.

In this work, we are interested in special models such that the states $%
|\chi \rangle $ \textit{exactly} span the ground-state subspace in the
thermodynamic limit. The specifications about the thermodynamic limit comes
from the fact that every state for $N$ sites can be written by means of a
matrix product ansatz given in Eq. (\ref{eq:MPS}) by taking $D>d^{\frac{N}{2}%
}$. However, we seek for models for which an exact renormalization can be
performed for arbitrarily long chains. In other words, the ground states of
these models can be solved rigorously through the real-space
renormalization, and the truncation induced by the Kraus operators does not
harm.

Practically, since $h_{i}\geq 0$, this search can be achieved if the Kraus
operators for each spin can be adjusted step by step in the renormalization
group to fulfill
\begin{equation}
\mathrm{Tr}(\rho _{i}^{\chi }h_{i})=0,\text{ \ }\forall i=1\cdots N\text{
and }\forall \chi =1\cdots D_{N}  \label{eq:GSkeeping}
\end{equation}%
where $\rho _{i}^{\chi }=\mathrm{Tr}_{\mathrm{env}}[|\chi \rangle \langle
\chi |]$ is the reduced density matrix for $k$ spins. The above condition
leads to $H|\chi \rangle =0$, which means that the vectors $|\chi \rangle $
are the ground states of $H$, because $H\geq 0$. Such Hamiltonians are
called \textit{frustration-free Hamiltonians} since their ground states
minimize energy locally. For instance, it is well-known that the
ferromagnetic Heisenberg chain is a typical frustration-free model in which
all the spins tends to align in parallel to gain energy. Recently, the
frustration-free Hamiltonians have been reformulated as quantum $k$-SAT
problems and attract considerable interests in quantum information community
\cite{Bravyi-2006,Shor-2010,Beaudrap-2010,Laumann-2010}.
% ********************************
% FIG: RSR
% ********************************
\begin{figure}%
    \centering
    \subfloat[][]{\label{fig:chap3_renorma}%
        \includegraphics[width=0.45\textwidth]{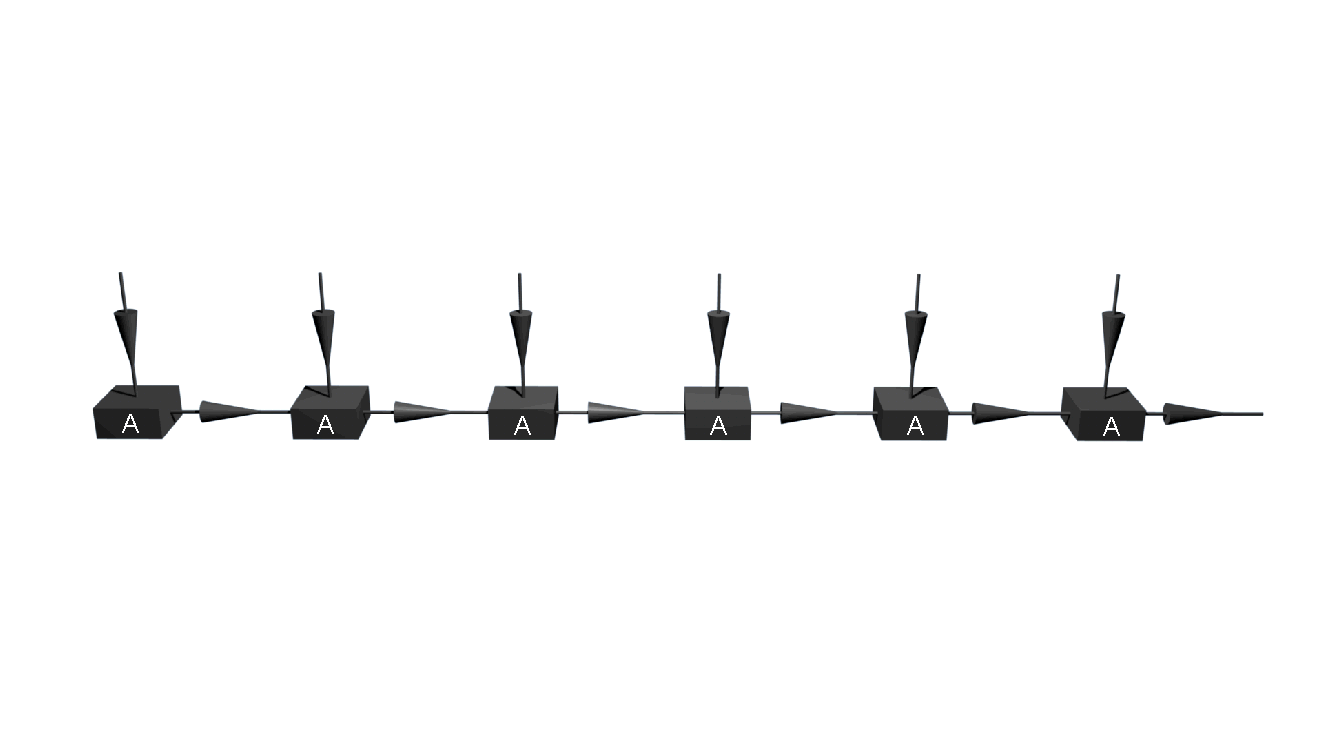}}%
    \qquad\quad
    \subfloat[][]{\label{fig:chap3_MPS-PBC}%
        \includegraphics[width=0.45\textwidth]{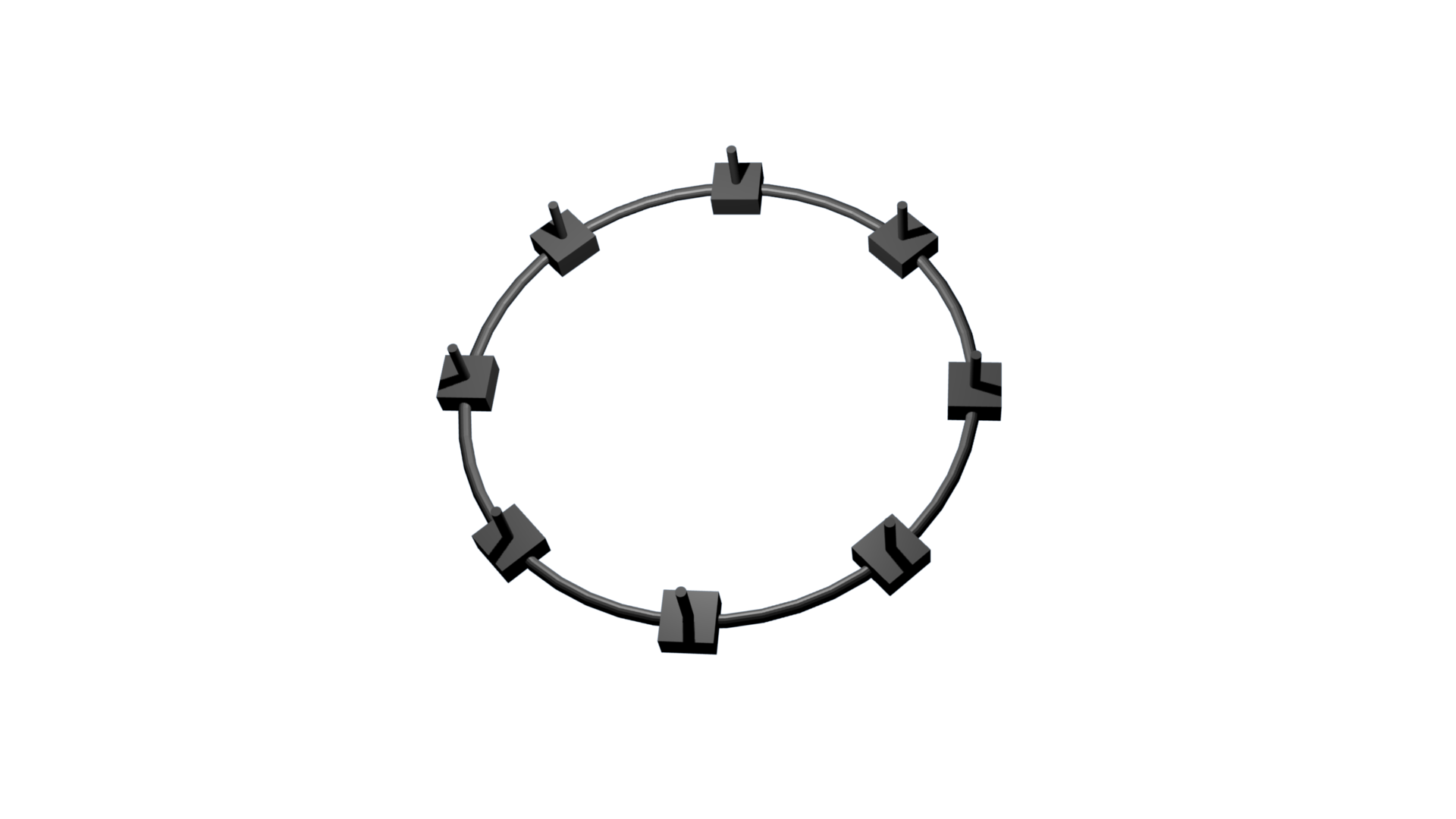}}
     \caption[Real space renormalization]{%
             \textbf{Real space renormalization}. (a) The real-space renormalization group yields matrix product states. (b) In periodic boundary condition, the translationally invariant MPS are constructed from the fixed point Kraus operators.}%
    \label{fig:chap3_real_space_renormalization}%
\end{figure}

For generic models, this renormalization procedure will terminate after
blocking a number of spins due to intrinsic frustrations. To find the
exactly solvable model, the first possibility is that the renormalization
group reaches a fixed point. Then, the ground state of the Hamiltonian in
periodic boundary condition can be written as a translationally invariant
MPS (See Fig. 1b)%
\begin{equation*}
|\Psi \rangle =\sum_{M_{1}\ldots M_{N}}\mathrm{Tr}(A^{[M_{1}]}A^{[M_{2}]}%
\cdots A^{[M_{N}]})|M_{1},M_{2}\cdots M_{N}\rangle
\end{equation*}%
where the Kraus operators $A^{[M]}$ are the converged $D\times D$ matrices
at the fixed point. We discuss these fixed point MPS solutions in Sec. III
B. Another possibility is that, for some models, the number of states
\textrm{dim}$\mathcal{H}_{k}$\ that we should keep,\textrm{\ }increases when
gathering more spins. Even though there is no renormalization fixed point,
we find that it is still possible to obtain the ground states exactly if
\textrm{dim}$\mathcal{H}_{k}$ increases in a \textit{controllable} way. We
illustrate this point in Sec. III C, when discussing the partially
ferromagnetic states.

\section{Quantum spin chains with SU(2) symmetry}

In this section,\textit{\ }we adapt the real-space exact renormalization to
SU(2) symmetric quantum spin chains with nearest-neighbor interactions.
Therefore, let us start by explaining some details about the SU(2) symmetric
Hamiltonians. The most general SU(2) symmetric translationally invariant
spin-$S$ Hamiltonian with nearest-neighbor interactions can be expressed as%
\begin{equation}
H=\sum_{i}\sum_{n=1}^{2S}a_{n}(\vec{S}_{i}\cdot \vec{S}_{i+1})^{n}+a_{0}%
\mathbbm{1}.  \label{eq:hamiltonian1}
\end{equation}%
The study of these SU(2) symmetric models has a long history in condensed
matter physics. It was known that some of these models can be solved by
Bethe Ansatz method and such models are fully classified by solutions of
Yang-Baxter equations \cite{Kennedy-1992}.

We want to identify the frustration-free models in Eq. (\ref{eq:hamiltonian1})
and find their ground states through real-space exact renormalization.
However, it is convenient to use projectors instead of spin operators, so we
use the transformation%
\begin{equation*}
(\vec{S}_{i}\cdot \vec{S}_{i+1})^{n}=\sum_{S_{T}=0}^{2S}[\frac{1}{2}%
S_{T}(S_{T}+1)-S(S+1)]^{n}P_{S_{T}}(i,i+1)
\end{equation*}%
where $P_{S_{T}}$ is a projector onto total-spin $S_{T}$ states of the two
spins. By shifting the local energy levels, we can always rewrite the
Hamiltonian (\ref{eq:hamiltonian1})\ as a sum of projectors
\begin{equation}
H=\sum_{i}\sum_{S_{T}\in \mathcal{K}}J_{S_{T}}P_{S_{T}}(i,i+1)
\label{eq:hamiltonian2}
\end{equation}%
with coupling constants $J_{S_{T}}>0$ and $\mathcal{K}\subseteq \lbrack
0,2S] $ is a set specifying the choice of projector(s) as local
interactions. Since the local interactions in Eq. (\ref{eq:hamiltonian2}) are a
sum of projectors, we have $H\geq 0$.

Let us remark that, as the physical representation is irreducible and we
restrict to nearest-neighbor interactions, the exact value of $J_{S_{T}}$ is
not important whenever the Hamiltonian is frustration-free.

From the projector Hamiltonian (\ref{eq:hamiltonian2}), it is still not
clear how to properly choose, if possible, the set $\mathcal{K}$ to make the
Hamiltonian frustration-free. However, as we restrict ourselves to
frustration-free models with two neighboring spins, we can provide a
complete list by taking advantage of the renormalization group.

\subsection{Exact renormalization with SU(2) symmetry}

In this subsection, we explain how to make use of the SU(2) symmetry in the
exact renormalization scheme. This particularizes the real-space
renormalization in Eq. (\ref{eq:isometry}) such that both the input and
output states form representations of the symmetry group, which ensures the
symmetry is preserved in each renormalization step. The method shown here is
a three-step process.

Equation (\ref{eq:isometry}) can be promoted to an SU(2) adapted basis \cite%
{Dukelsky-1998,McCulloch-2002,Vidal-2009}%
\begin{equation}
|j_{b}t_{b}m_{b}\rangle
=\sum_{j_{a}t_{a}m_{a}}%
\sum_{M}A_{j_{a}t_{a}m_{a},j_{b}t_{b}m_{b}}^{[S,M]}|j_{a}t_{a}m_{a}\rangle
|SM\rangle  \label{eq:su2renormalization}
\end{equation}%
where the $j$'s denote the SU(2) representations (total-spin quantum
number), the $t$'s distinguish the degenerate states within the same $j$,
and the $m$'s are the magnetic quantum numbers associated with $j$.

The first step of the process consists of splitting the Kraus operators into
two terms by means of Wigner-Eckart theorem (See Fig. 2a) as
\begin{equation}
A_{j_{a}t_{a}m_{a},j_{b}t_{b}m_{b}}^{[S,M]}=T_{j_{a}t_{a},j_{b}t_{b}}\langle
j_{a}m_{a},SM|j_{b}m_{b}\rangle  \label{eq:su2isometry}
\end{equation}%
where the indices $j_{a}t_{a},j_{b}t_{b}$ keep track of the representations
of the input and output states. The first term is a real matrix $T$ denoting
the weights of different input states in each output states. We call this
matrix \textquotedblleft weight matrix".\ Let us remark that the weight
matrix does not depend on the magnetic quantum numbers. The second term is
the Clebsch-Gordan coefficient $\langle j_{a}m_{a},SM|j_{b}m_{b}\rangle $,\
corresponding to the representation fusion $j_{a}\otimes S\rightarrow j_{b}$%
. To ensure that the output states always form an orthonormal basis, the
weight matrix must fulfill%
\begin{equation}
T_{j_{a}t_{a},j_{b}t_{b}}=0\text{ \textrm{unless} }|j_{a}-S|\leq j_{b}\leq
j_{a}+S  \label{eq:constraint}
\end{equation}%
\begin{equation}
\sum_{j_{a}t_{a}}T_{j_{a}t_{a},j_{b}t_{b}}T_{j_{a}t_{a},j_{b}t_{b}^{\prime
}}=\delta _{t_{b},t_{b}^{\prime }}  \label{eq:Tmat}
\end{equation}%
for every $j_{b}$. The first constraint is related to SU(2) fusion rules.
The second constraint means the columns of $T_{j_{a}t_{a},j_{b}t_{b}}$
corresponding to the same $j_{b}$ but different $t_{b}$, are orthonormal
vectors, which guarantees the isometry condition $\sum_{M}A^{[S,M]\dagger
}A^{[S,M]}=\mathbbm{1}$ for the Kraus operators.

The advantage of this representation for the Kraus operators is that it
allows us to design an elegant way to perform the exact renormalization
group, which is the second step of the method. Let us consider two
neighboring spins (See Fig. 2b). The renormalization process consists of two
sequential representation fusions $j_{a}\otimes S\rightarrow j_{b}$ and $%
j_{b}\otimes S\rightarrow j_{c}$. As a result, we obtain the orthonormal
basis%
\begin{eqnarray}
|j_{c}t_{c}m_{c}\rangle
&=&\sum_{M_{1}M_{2}}\sum_{j_{a}t_{a}m_{a}}%
\sum_{j_{b}t_{b}m_{b}}T_{j_{a}t_{a},j_{b}t_{b}}  \notag \\
&&\times \langle j_{a}m_{a},SM_{1}|j_{b}m_{b}\rangle \langle
j_{b}m_{b},SM_{2}|j_{c}m_{c}\rangle   \notag \\
&&\times T_{j_{b}t_{b},j_{c}t_{c}}^{\prime }|j_{a}t_{a}m_{a}\rangle
|SM_{1}\rangle |SM_{2}\rangle .  \label{eq:fusion1}
\end{eqnarray}%
where the weight matrices $T$ and $T^{\prime }$ for these two spins can be
different in general. Alternately, the renormalization process Eq. (\ref%
{eq:fusion1}) can be done by fusion of the two physical spins to their
coupled representations $S\otimes S\rightarrow S_{T}$ first and then $%
j_{a}\otimes S_{T}\rightarrow j_{c}$. In the latter fusion sequence, we
obtain the same basis%
\begin{eqnarray}
|j_{c}t_{c}m_{c}\rangle
&=&\sum_{j_{a}t_{a}m_{a}}\sum_{S_{T}M_{T}}R_{j_{a}t_{a},j_{c}t_{c}}^{S_{T}}%
\langle j_{a}m_{a},S_{T}M_{T}|j_{c}m_{c}\rangle   \notag \\
&&\times |j_{a}t_{a}m_{a}\rangle |S_{T}M_{T}\rangle   \label{eq:fusion2}
\end{eqnarray}%
where $|S_{T}M_{T}\rangle $ is the coupled basis of two physical spins. The
two different fusion channels are unitarily related by the recoupling $F$%
-symbol (See Fig. 2c) defined by $F_{SS_{T}j_{c}}^{j_{a}Sj_{b}}=\langle
j_{b}(j_{a}S),S;j_{c}m_{c}|j_{a},S_{T}(SS);j_{c}m_{c}\rangle $. By using
Wigner's $6$-$j$ symbol, this $F$-symbol can be expressed as \cite%
{Brink-Satchler}%
\begin{equation*}
F_{SS_{T}j_{c}}^{j_{a}Sj_{b}}=(-1)^{j_{a}+j_{c}}\sqrt{(2j_{b}+1)(2S_{T}+1)}%
\begin{Bmatrix}
j_{a} & S & j_{b} \\
S & j_{c} & S_{T}%
\end{Bmatrix}%
.
\end{equation*}%
By substituting this in Eq. (\ref{eq:fusion1}) and comparing with Eq. (\ref%
{eq:fusion2}), we obtain%
\begin{equation}
R_{j_{a}t_{a},j_{c}t_{c}}^{S_{T}}=%
\sum_{j_{b}t_{b}}T_{j_{a}t_{a},j_{b}t_{b}}F_{SS_{T}j_{c}}^{j_{a}Sj_{b}}T_{j_{b}t_{b},j_{c}t_{c}}^{\prime }.
\label{eq:weight}
\end{equation}%
According to Eq. (\ref{eq:fusion2}), the output states $|j_{c}t_{c}m_{c}%
\rangle $ only keep the local ground states of the Hamiltonian (\ref%
{eq:hamiltonian2}) if%
\begin{equation}
R_{j_{a}t_{a},j_{c}t_{c}}^{S_{T}}=0\text{ \ }\forall j_{a},t_{a}\text{
\textrm{and} }S_{T}\in \mathcal{K}  \label{eq:renormalization}
\end{equation}%
This equation relates the weight matrices of two spins in Eq. (\ref{eq:weight})
and plays an important role in our exact renormalization group method.
% ********************************
% FIG: RSR SU(2)
% ********************************
\begin{figure}%
    \centering
    \subfloat[][]{\label{fig:chap3_figHH3b}%
        \includegraphics[width=0.45\textwidth]{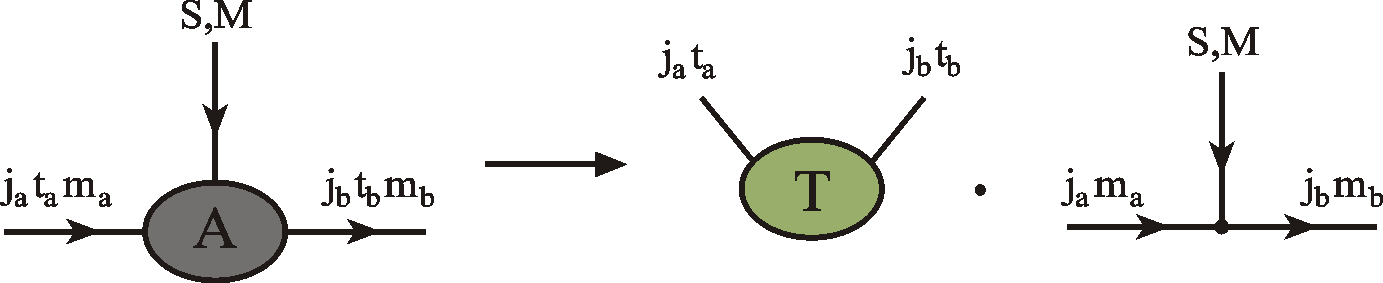}}%
    \vspace{0.5cm} \qquad\quad
    \subfloat[][]{\label{fig:chap3_figHH4}%
        \includegraphics[width=0.45\textwidth]{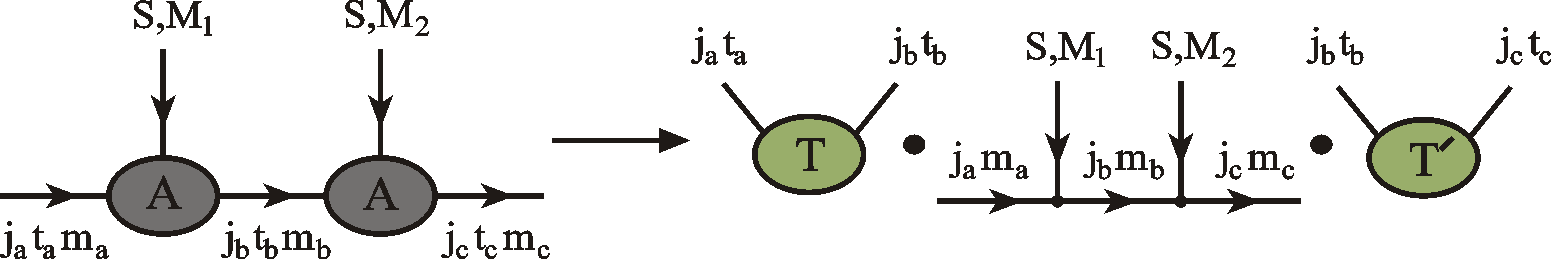}}%
    \vspace{0.5cm}\qquad\quad
    \subfloat[][]{\label{fig:chap3_figHH5}%
        \includegraphics[width=0.45\textwidth]{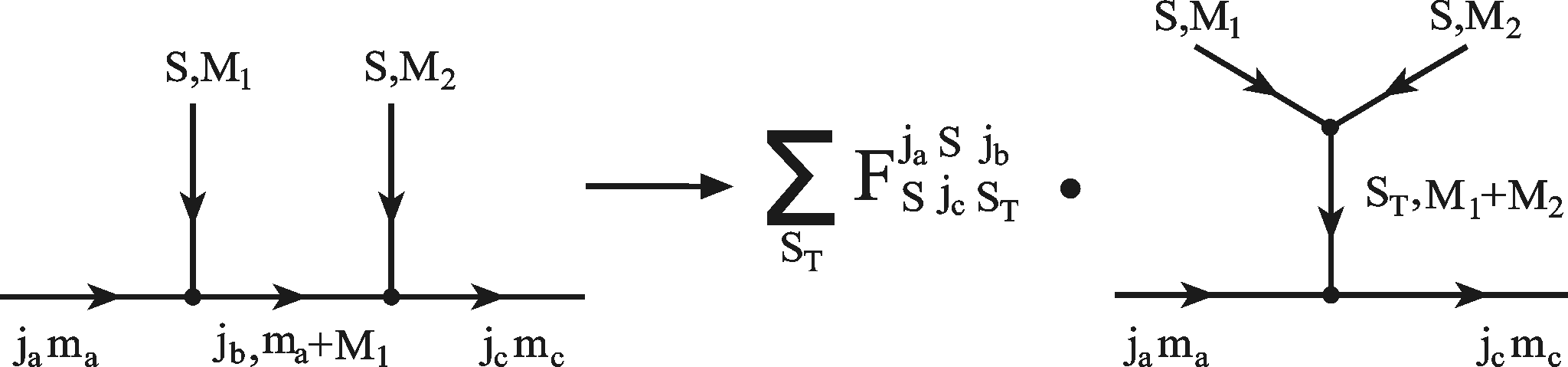}}
    \caption[Real space renormalization with $SU(2)$]{%
        \textbf{Real space renormalization with $SU(2)$}. (a) An isometry $A^{[S,M]}$ is decomposed as a matrix $T$ and a fusion of angular momentums. (b) The renormalization of two spins can be done in two successive steps. (c) The same input and output states with different intermediate fusion channels for two spins can be related by the $F$-symbol.}%
    \label{fig:chap3_fig2}%
\end{figure}

The third step of our method is to use Eq. (\ref{eq:renormalization}) to
carry out the renormalization group process for the whole spin chain.
Hereafter we use $T^{[i]}$ to denote the weight matrix at site $i$. Let us
start the renormalization from the first two spins. By taking the first
physical spin $S$ as the input representation, we have the initial condition
$T_{j_{1}}^{[1]}=1$ with $j_{1}=S$. According to Eq. (\ref{eq:weight}) and
Eq. (\ref{eq:renormalization}), we obtain $T_{j_{1},j_{2}}^{[2]}=1$. The
output representation $j_{2}\in \mathcal{\bar{K}}$, where $\mathcal{\bar{K}}$
is the orthogonal complement of $\mathcal{K}$. This simple test verifies the
output states are the zero-energy local ground states of the projector
Hamiltonian (\ref{eq:hamiltonian2}).

The renormalization group follows naturally as $T^{[2]}\rightarrow
T^{[3]}\rightarrow \cdots $ under the renormalization condition Eq. (\ref%
{eq:renormalization}) and the constraints Eq. (\ref{eq:constraint}) and Eq. (\ref%
{eq:Tmat}). Let us describe how to deal with these requirements
simultaneously. Let us suppose that we already know the weight matrix $T^{[i-1]}$ and
the goal is to calculate $T^{[i]}$. After taking the square of Eq. (\ref%
{eq:renormalization}) and summing over $j_{a},t_{a}$ and $S_{T}\in \mathcal{K%
}$, we obtain
\begin{equation}
\sum_{j_{b}^{\prime }t_{b}^{\prime }}\sum_{j_{b}t_{b}}T_{j_{b}^{\prime
}t_{b}^{\prime },j_{c}t_{c}}^{[i]}\mathcal{M}_{j_{b}^{\prime }t_{b}^{\prime
},j_{b}t_{b}}^{[i]j_{c}}T_{j_{b}t_{b},j_{c}t_{c}}^{[i]}=0
\end{equation}%
where the positive semidefinite real Hermitian\ matrix $\mathcal{M}%
^{[i]j_{c}}$ is given by
\begin{equation*}
\mathcal{M}_{j_{b}^{\prime }t_{b}^{\prime
},j_{b}t_{b}}^{[i]j_{c}}=\sum_{S_{T}\in \mathcal{K}%
}\sum_{j_{a}t_{a}}T_{j_{a}t_{a},j_{b}^{\prime }t_{b}^{\prime
}}^{[i-1]}F_{SS_{T}j_{c}}^{j_{a}Sj_{b}^{\prime
}}F_{SS_{T}j_{c}}^{j_{a}Sj_{b}}T_{j_{a}t_{a},j_{b}t_{b}}^{[i-1]}.
\end{equation*}%
For every possible $j_{c}$ from $j_{b}\otimes S$, we calculate the kernel of
$\mathcal{M}^{[i]j_{c}}$, which give us the weight matrix $T^{[i]}$. If $%
\mathcal{M}^{[i]j_{c}}$ does not have kernel vectors satisfying Eq. (\ref%
{eq:constraint}), the corresponding output representation $j_{c}$ must be
discarded. If the kernel of $\mathcal{M}^{[i]j_{c}}$ has dimension larger
than $1$, the index $t_{c}$ is used to tag the orthonormal kernel vectors
for such $j_{c}$. Thus, the kernel vectors of $\mathcal{M}^{[i]j_{c}}$
constitute the columns of $T^{[i]}$ and the column indices $j_{c},t_{c}$ of $%
T_{j_{b}t_{b},j_{c}t_{c}}^{[i]}$ denote the output representations.\ One can
straightforwardly show that the resulting weight matrix $T^{[i]}$ satisfies
the renormalization condition Eq. (\ref{eq:renormalization}) and the orthonormal
constraint Eq. (\ref{eq:Tmat}), because $\mathcal{M}^{[i]j_{c}}$ is positive
semidefinite and Hermitian.

\subsection{Matrix product states}

In this subsection, we discuss the models which have a renormalization fixed
point and then, MPS as ground states. In our present exact renormalization
scheme, the renormalization fixed point means that the output
representations does not change when adding new spins and $T^{[i]}$
converges to a site-independent matrix.

Let us start by introducing two relevant concepts about MPS -- injectivity
and symmetry. We begin with the definition of injectivity \cite{Sanz-2010}

\begin{definition}[Injectivity]
\label{def:chap1_injectivity} Let $|\alpha \rangle \in \mathbb{C}^{D}$ be an
orthonormal basis and $\{A^{[M]}\}_{M=1}^{d}$ the $D\times D$ Kraus
operators defining a translationally invariant MPS. And let us consider the $%
D^{2}$ states for $L$ sites defined as
\begin{equation}
|\psi _{\alpha \beta }^{(L)}\rangle =\sum_{M_{1}\cdots M_{L}}\langle \alpha
|A^{[M_{1}]}\cdots A^{[M_{L}]}|\beta \rangle |M_{1}\cdots M_{L}\rangle \label{eq:injectivity}
\end{equation}%
Then, we say that the MPS is \textit{injective} (see Fig. 3) if there exists a finite $L$
such that the vector space spanned by the vectors in Eq. (\ref{eq:injectivity}) has dimension $D^{2}$. In other words,
different boundary conditions turn into different states. The injectivity
length $L_{0}$ is defined by the minimal number of sites for which
injectivity is reached.
\end{definition}

% ********************************
% FIG: MPS-OBC
% ********************************
\begin{figure}%
    \centering %
        \includegraphics[width=0.45\textwidth]{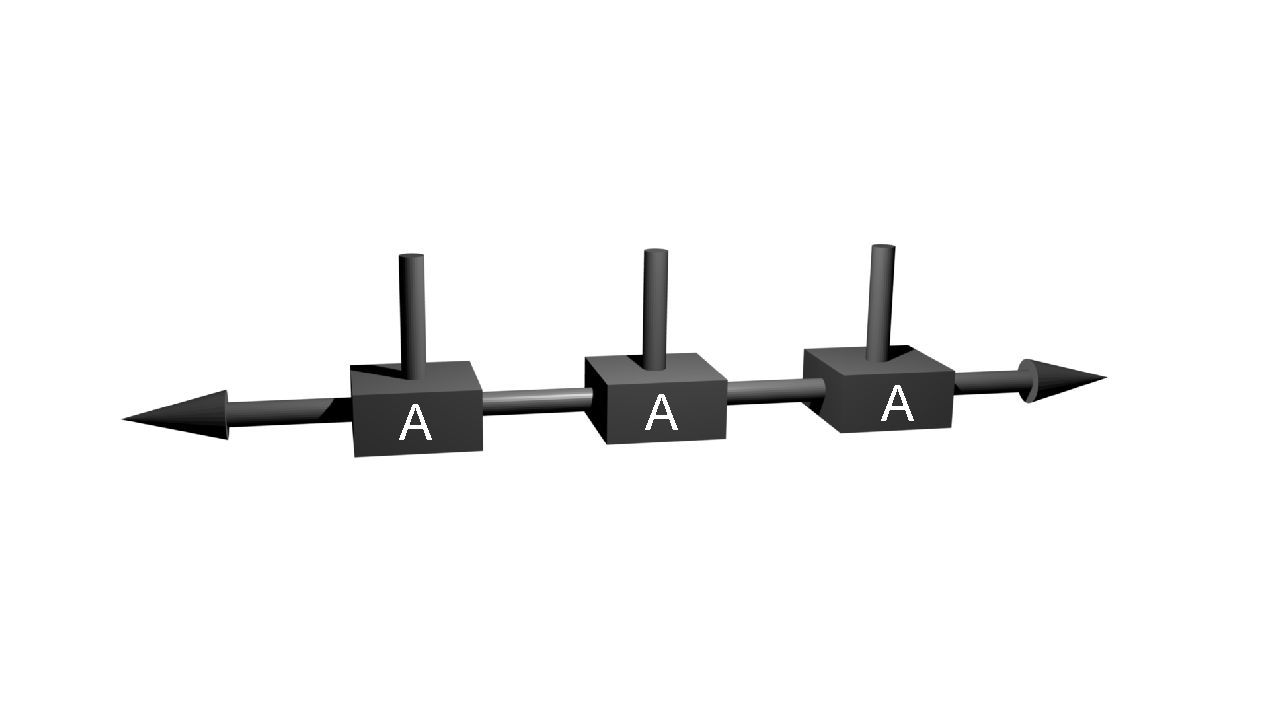}%
    \caption[Injectivity]{%
        \textbf{Injectivity}. A MPS is injective with injectivity length $L_0$ if for the MPS constructed for $L_0$ sites, different boundary conditions (linearly independent), represented by the cones in the figure, give rise to different states (linearly independent), and this does not happen for $L_0-1$ spins.}%
    \label{fig:chap1_injectivity}%
\end{figure}

The interest of this definition comes from Ref \cite{Fannes-1989,Cirac-2007}%
, where it is proven that injectivity is the necessary and sufficient
condition for the existence of a parent Hamiltonian which has the MPS as a
unique ground state with a non-trivial spectral gap above.

The other relevant result that we would like to recall here is the
construction of translationally invariant MPS which are locally invariant
under some symmetry group $G$. The following theorem provides the necessary
and sufficient conditions \cite{Sanz-2009} for that

\begin{theorem}[Symmetry]
Let $|\Psi \rangle \in (\mathbb{C}^{d})^{\otimes N}$ be translationally invariant MPS defined by the Kraus operators $\{A^{[M]}\}_{M=1}^{d}$, and let $u$ and $U$ be two representations of a finite or a compact Lie group $G$. Then, $|\Psi \rangle $ is invariant under $G$ in the sense of $u^{\otimes N}|\Psi \rangle =e^{i\theta N}|\Psi \rangle $ if and only if (see Fig. 4)
\begin{equation}
\sum_{M^{\prime }}u_{M,M^{\prime }}A^{[M^{\prime }]}=e^{i\theta
}UA^{[M]}U^{\dagger }
\end{equation}
\end{theorem}

% ********************************
% FIG: Symmetry condition
% ********************************
\begin{figure}%
    \centering
        \includegraphics[width=0.45\textwidth]{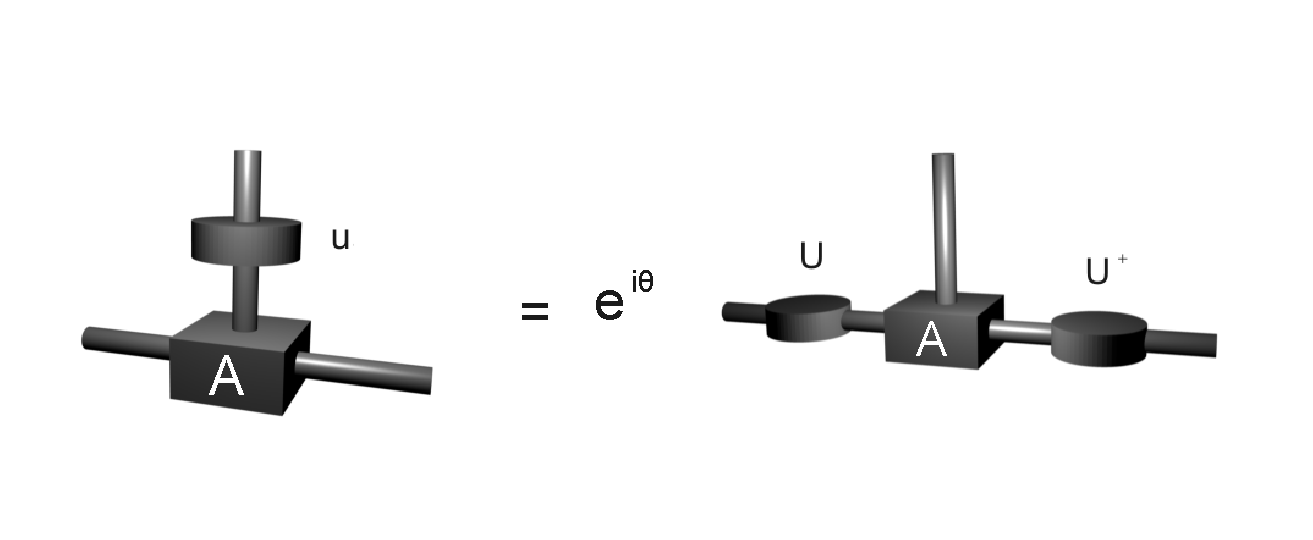}
    \caption[Symmetry conditions]{%
        \textbf{Symmetry}.
        The unitary $u$ applied on the physical level is reflected in the virtual level as pair of unitaries $U$, as shown in Ref. \cite{Sanz-2009}}%
    \label{fig:chap1_Krauscond}%
\end{figure}

Here we call that $u$ and $U$ are the physical and virtual spin
representations, respectively. Once these representations are fixed, the
Kraus operators can be constructed by means of the Clebsch-Gordan
coefficients together with a weight matrix. In our present SU(2) case, the
Kraus operators are exactly given by the decomposition in Eq. (\ref%
{eq:su2isometry}).

Let us now remind the previous results about the MPS solutions for 
Hamiltonian (\ref{eq:hamiltonian2}). The best-known models belong to the
AKLT family \cite{Affleck-1987}, which are defined by $\mathcal{K}%
=\{S+1,S+2,\ldots ,2S\}$ for integer-spin $S$. The MPS of spin-$S$ AKLT
model have a VBS picture with irreducible virtual spin-$S/2$ representation.
Another family of the models\ also have integer spin and the Hamiltonians
are defined by $\mathcal{K}=\{2,4,\ldots ,2S\}$ \cite{Tu-2008,Tu-2009},
which we call SO$(2S+1)$ symmetric family. For the $S=2$ model of this
family, the MPS have irreducible virtual spin-$3/2$ representations \cite%
{Tu-2008}, which is equivalent to the SO(5) symmetric MPS in a two-leg
electronic ladder \cite{Scalapino-1998}. For $S\geq 3$ cases, the properties
of the corresponding MPS are less clear, even though their explicit wave
functions were found \cite{Tu-2008}.

Now we turn to our results obtained by the exact renormalization group. For
the Hamiltonian (\ref{eq:hamiltonian2}), we check all possible $\mathcal{K}$\
\cite{Method}\ and then provide a complete list of fixed point MPS solutions
for $S\leq 4$, and a new solution for $S=5$. All these solutions are
integer-spin models \cite{Semi-integer}, which are summarized in Table \ref%
{tab:MPS}. For $S\leq 4$, we conclude that there is no solution other than
the above two families. For $S=5$, we find a new model, whose Hamiltonian is
given by $\mathcal{K}=\{3,7,8,9,10\}$ and the ground state has a VBS picture
with irreducible virtual spin-$3$ representations.

For the SO$(2S+1)$ family with $S\geq 3$, the exact renormalization group
provides us a more comprehensible physical picture, which can be viewed as
generalized VBS with \textit{reducible} virtual spin representations. In
Table \ref{tab:MPS}, we also listed the minimal number of blocked spins to
reach the fixed point representations. Since all these MPS are injective,
this length scale is actually the injectivity length \cite{Sanz-2010}.

Let us explain these results with an explicit example in the SO$(2S+1)$
family: the spin-$3$ model with $\mathcal{K}=\{2,4,6\}$. Through the exact
renormalization group, we can observe that the output states reach the
fixed point representation $0\oplus 0\oplus 1\oplus 2\oplus 3\oplus 3\oplus
3\oplus 4\oplus 5\oplus 6$ after blocking $6$ spins. To obtain the MPS, we
do not really need to calculate the fixed point Kraus operators by the
renormalization group. According to Theorem 2, the fixed point
representations allow us to construct this MPS directly. For the present
example, the fixed point representations give an important hint that the MPS
has a VBS picture (See Fig. 5a) with SU(2) \textit{reducible} virtual spin
representation $0\oplus 3$, which is quite different from the traditional
VBS states with irreducible virtual spin representations, like AKLT states
\cite{Affleck-1987} or their extensions \cite{Fannes-1989}.

With a chain beyond the injectivity length $L_{0}=6$, the tensor product of
two $0\oplus 3$ representations at the two boundaries yields the observed
fixed point representation in the renormalization group. For open boundary
conditions, in thermodynamic limit, the unpaired representations $0\oplus 3$
at the two edges are asymptotically free and become well-defined \textit{%
edge states}. For periodic boundary conditions, all virtual spin
representations are contracted into SU(2) singlets with neighboring sites
and therefore the MPS is a global spin singlet.
% ********************************
% FIG: RSR
% ********************************
\begin{figure}%
    \centering
    \subfloat[][]{\label{fig:chap3_figHH1}%
        \includegraphics[width=0.45\textwidth]{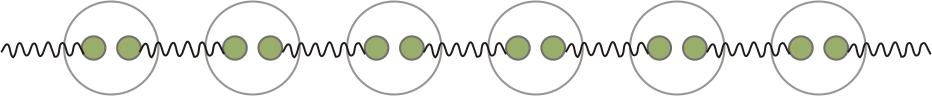}}%
    \vspace{0.5cm}\qquad\quad
    \subfloat[][]{\label{fig:chap3_figHH2}%
        \includegraphics[width=0.45\textwidth]{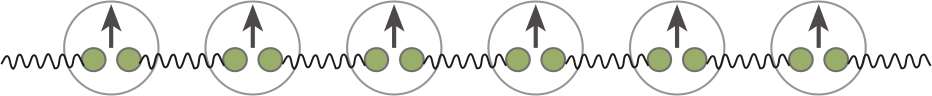}}
    \caption[Results of the technique]{%
        \textbf{Ground-state physical picture}. (a) The fixed point type MPS solutions have a VBS picture. Each dot denotes a virtual spin representation. The wavy lines represent the valence-bond singlets between virtual spins and the circles indicate the projection of two virtual spins onto physical spin representations. In the exact renormalization calculations, the fixed point representations are from the tensor product of two virtual spins (edge states). (b) The partially ferromagnetic states have a magnetization plateau. The arrows denote a fully polarized virtual spin-($S-1$) in a spin-$S$ partially ferromagnetic state.}%
    \label{fig:fig3}%
\end{figure}

The renormalization group analysis has also been carried out for other models in SO($2S+1$) family. From Table \ref{tab:MPS}, one can see that, for $S\geq 3$, their matrix product ground states have reducible virtual spin representations, which directly correspond to the edge states in an open chain. This provides more complete understanding of these systems. For all MPS in Tab. \ref{tab:MPS}, we present their explicit Kraus operators in
Appendix A.
% ********************************
% FIG: RSR
% ********************************
\begin{table}
\begin{center}
\begin{tabular}[c]{c | c c c}
\textbf{Spin} & \textbf{Set $\mathcal{K}$} & \textbf{Virtual spin} & \textbf{$L_0$}\\
[.2em] \hline \hline\\[-.9em]
1   & $\{2\}$             & $1/2$       & 2\\
2   & $\{3,4\}$           & $1$         & 2\\
2   & $\{2,4\}$           & $3/2$       & 4\\
3   & $\{4,5,6\}$         & $3/2$       & 2\\
3   & $\{2,4,6\}$         & $0\oplus 3$ & 6\\
4   & $\{5,6,7,8\}$       & $2$         & 2\\
4   & $\{2,4,6,8\}$       & $2\oplus 5$ & 8\\
5   & $\{6,7,8,9,10\}$    & $5/2$       & 2\\
5   & $\{2,4,6,8,10\}$    & $5/2\oplus 9/2 \oplus 15/2$         & 10\\
5   & $\{3,7,8,9,10\}$    & $3$         & 4
\end{tabular}
\caption{Models with $SU(2)$--invariance, nearest--neighbour interactions and matrix product ground states. $L_{0}$ is the
injectivity length.}%
\label{tab:MPS}%
\end{center}
\end{table}

Let us make a remark about these exactly solvable models. All their
fixed point MPS ground states have exponentially decaying correlations and
there is an energy gap above the ground states, since they are injective. However, the different
virtual spin representations (edge states) show that these MPSs belong to
different quantum phases of matter. Therefore, once a new Hamiltonian $%
H=(1-x)H_{1}+xH_{2}$ is constructed from two solvable models $H_{1}$ and $%
H_{2}$ in Table \ref{tab:MPS} with the same spin $S$, at least one quantum
phase transition is expected to occur when tuning $x$ from $0$ to $1$. Since
both MPS ground states for $H_{1}$ and $H_{2}$ preserve SU(2) symmetry, the
local order parameter description breaks down and unconventional quantum
phase transitions may emerge. Very recently, this idea has been exploited to
study the possibility of a topological quantum phase transition in an $S=2$
chain \cite{Zheng-Zang-2010}.

\subsection{Partially ferromagnetic states}

In this subsection, we discuss another class of models which do not have
renormalization fixed points but still can be solved exactly. The ground
states of these models are partially ferromagnetic states.

This family includes both semi-integer spin models and integer-spin models.
The Hamiltonian is defined by $\mathcal{K}=\{0,1,\ldots 2S-4,2S\}$ and the
physical spin $S\geq 5/2$. Their ground states are partially ferromagnetic
states with a magnetization plateau $\langle S_{i}^{z}\rangle =S-1$. We also
have found a physical picture (See Fig. 5b) for these states with partial
magnetization: We prepare a spin-$1$ AKLT-type VBS state with virtual spin-$%
1/2$ and a spin-$(S-1)$ maximally polarized ferromagnetic state. In each
site, we recover the physical spin-$S$ Hilbert space by $(S-1)\otimes
1\rightarrow S$, which is achieved by applying local projections.

Let us consider a typical example -- the spin-$5/2$ model with $\mathcal{K}%
=\{0,1,5\}$. For a block of $N_{0}$ spins, the AKLT part contributes
representations $0\oplus 1$ and the polarized ferromagnetic part contributes
representation $3N_{0}/2$. Thus, the total spin of the $N_{0}$-spin\ block
is given by the tensor product of representations from these two parts%
\begin{equation}
(0\oplus 1)\otimes \frac{3N_{0}}{2}=(\frac{3N_{0}}{2}-1)\oplus \frac{3N_{0}}{%
2}\oplus \frac{3N_{0}}{2}\oplus (\frac{3N_{0}}{2}+1)  \label{eq:partialFM}
\end{equation}%
For two spins ($N_{0}=2$), the allowed representations are $2\oplus 3\oplus
3\oplus 4$ and can not reach $\mathcal{K}=\{0,1,5\}$, which means that the
partially ferromagnetic state is the zero-energy ground state of the
projector Hamiltonian. In the exact renormalization process, we found that
the four output representations in Eq. (\ref{eq:partialFM}) are the only
output representations for $N_{0}\geq 6$. By adding one additional spin,
the total spin of the four representations is increased by $3/2$. These
observations actually strongly suggest the partially ferromagnetic picture
of the ground state.

One may ask why this class starts with $S=5/2$ rather than $S=2$. The reason
is the following: For spin-$2$ model $\mathcal{K}=\{0,4\}$, the
renormalization group shows that the number of output representations does
not saturate, which means that the partially ferromagnetic state is not the
only ground state of the Hamiltonian.

Compared to the fixed point MPS solutions in Sec. III B, the partially
ferromagnetic states have a long range order and thus break the SU(2)
symmetry. According to Goldstone theorem, we expect gapless spin wave
excitations above the ground state, which is quite different from the gapped
fixed point MPS with exponentially decaying correlations.

\section{Conclusions and Perspectives}

We have introduced a real-space exact renormalization group adapted to the
SU(2) symmetry, which is well suited for finding exactly solvable quantum
spin Hamiltonians with nearest-neighbor interactions.

The list of solutions can be divided into two classes according to the
renormalization group behavior. In the first class, the models are quantum
integer-spin chains with renormalization fixed points and matrix product
ground states. For $S\leq 4$, we show that the AKLT family and the SO$(2S+1)$
family exhaust all possible solutions. In the SO$(2S+1)$ family, the
renormalization group provides a natural explanation for the edge states of
the MPS by providing a generalized VBS picture with reducible virtual spin
representation. Furthermore, we obtain a new solvable model for $S=5$ beyond
the existing families. In the second class, the models have partially
ferromagnetic ground states with a magnetization plateau. This solvable
family exists for $S\geq 5/2$ and contains both integer spin and
semi-integer spin models. The partially ferromagnetic ground states have
gapless spin-wave excitations, which are quite different from the gapped MPS
in the first class.

Beyond the present work, it would be quite interesting to generalize the
method to spin chains beyond nearest-neighbor interactions and models in
higher dimensions, especially adapted to PEPS formalism. Furthermore, the
method may be used to explain an open question by \"{O}stlund and Rommer
\cite{Ostlund-1995} about which representations must be introduced in the
renormalization group with SU(2) symmetry.

Finally, we also expect a natural extension of the present exact
renormalization formalism to quantum spin chains with other symmetry groups
\cite{Rachel-2007}.

\begin{acknowledgments}
The authors would like to thank J. Ignacio Cirac, Miguel Aguado, and Stephan Rachel for the
very fruitful discussions and A. Nogueira for his invaluable technical assistance. M. Sanz thanks the support of the QCCC Program of
the EliteNetzWerk Bayern.
\end{acknowledgments}

\appendix{}

\section{Kraus operators of the matrix product states}

In this Appendix, we explicitly present the Kraus operators needed for the
definition of the MPS in Table \ref{tab:MPS}. As we mentioned, the Kraus
operators with SU(2) symmetry are parametrized by Eq. (\ref{eq:su2isometry}%
), which requires both the set $\mathcal{V}$ containing the SU(2) virtual
spin representations and the weight matrix $T$.

For irreducible virtual spin representations, the set $\mathcal{V}$ contains
a single representation $j_{a}$ and therefore $T=1$. In this case, the Kraus
operators are simply the Clebsch-Gordan coefficients%
\begin{equation}
A_{j_{a}m_{a},j_{a}m_{b}}^{[S,M]}=\langle j_{a}m_{a},SM|j_{a}m_{b}\rangle .
\end{equation}

For reducible virtual spin representations, the set $\mathcal{V}$ has
multiple SU(2) representations and the weight matrix $T$ is necessary. The
Kraus operators are given by%
\begin{equation}
A_{j_{a}m_{a},j_{b}m_{b}}^{[S,M]}=T_{j_{a},j_{b}}\langle
j_{a}m_{a},SM|j_{b}m_{b}\rangle .
\end{equation}%
where the index $t$ is suppressed because no degeneracy occurs in $\mathcal{V%
}$ for our models. We use a convention to define the matrix $T$ such that
the row and the column indices $j_{a},j_{b}$ are arranged in an incremental\
order. For instance, the $S=3$ model with $\mathcal{K}=\{2,4,6\}$ has
virtual representation $0\oplus 3$ and%
\begin{equation}
T=%
\begin{pmatrix}
T_{0,0} & T_{0,3} \\
T_{3,0} & T_{3,3}%
\end{pmatrix}%
=%
\begin{pmatrix}
0 & \frac{-1}{\sqrt{7}} \\
1 & \sqrt{\frac{6}{7}}%
\end{pmatrix}%
.
\end{equation}%
For the $S=4$ model with $\mathcal{K}=\{2,4,6,8\}$, we have virtual
representation $2\oplus 5$ and%
\begin{equation}
T=%
\begin{pmatrix}
\frac{1}{3}\sqrt{\frac{7}{2}} & \frac{1}{3}\sqrt{\frac{5}{2}} \\
\frac{-1}{3}\sqrt{\frac{11}{2}} & \frac{1}{3}\sqrt{\frac{13}{2}}%
\end{pmatrix}%
.
\end{equation}%
For the $S=5$ model with $\mathcal{K}=\{2,4,6,8,10\}$, we have virtual
representation $5/2\oplus 9/2\oplus 15/2$ and%
\begin{equation}
T=%
\begin{pmatrix}
\frac{1}{11}\sqrt{\frac{21}{2}} & \frac{-3}{\sqrt{22}} & \frac{-1}{11}\sqrt{%
\frac{21}{2}} \\
-\sqrt{\frac{15}{22}} & \sqrt{\frac{3}{26}} & -\sqrt{\frac{85}{286}} \\
\frac{2\sqrt{7}}{11} & 2\sqrt{\frac{17}{143}} & \frac{1}{11}\sqrt{\frac{969}{%
13}}%
\end{pmatrix}%
.
\end{equation}%
It is straightforward to show that these solutions satisfy Eqs. (\ref%
{eq:constraint}), (\ref{eq:Tmat}), and (\ref{eq:renormalization}).


\begin{thebibliography}{99}
\bibitem{Cirac-review} for recent reviews, see J. I. Cirac and F.
Verstraete, J. Phys. A \textbf{42}, 504004 (2009); F. Verstraete, V. Murg,
and J. I. Cirac, Adv. Phys. \textbf{57}, 143 (2008).

\bibitem{Verstraete-2006} F. Verstraete and J. I. Cirac, Phys. Rev. B
\textbf{73}, 094423 (2006).

\bibitem{Hastings-2007} M. B. Hastings, J. Stat. Mech. (2007), P08024.

\bibitem{Eisert-2010} J. Eisert, M. Cramer, and M. B. Plenio, Rev. Mod.
Phys. \textbf{82}, 277 (2010).

\bibitem{Fannes-1989} M. Fannes, B. Nachtergaele, and R. F. Werner, Commun.
Math. Phys. \textbf{144}, 443 (1992).

\bibitem{Klumper-1991} A. Kl\"{u}mper, A. Schadschneider, and J. Zittartz,
J. Phys. A \textbf{24}, L955 (1991); Z. Phys. B: Condens. Matter \textbf{87}%
, 281 (1992).

\bibitem{Cirac-2007} D. P\'{e}rez-Garc\'{\i}a, F. Verstraete, M. M. Wolf,
and J. I. Cirac, Quantum Inf. Comput. \textbf{7}, 401 (2007).

\bibitem{White-DMRG} S. R. White, Phys. Rev. Lett. \textbf{69}, 2863 (1992).

\bibitem{Schollwock-review} U. Schollw\"{o}ck, Rev. Mod. Phys. \textbf{77},
259 (2005).

\bibitem{Ostlund-1995} S. \"{O}stlund and S. Rommer, Phys. Rev. Lett.
\textbf{75}, 3537 (1995).

\bibitem{Verstraet-Cirac-DMRG} F. Verstraete, D. Porras, and J. I. Cirac,
Phys. Rev. Lett. \textbf{93}, 227205 (2004).

\bibitem{Vidal-2004} G. Vidal, Phys. Rev. Lett. \textbf{93}, 040502 (2004).

\bibitem{Cirac-Sierra-2010} J. I. Cirac and G. Sierra, Phys. Rev. B \textbf{%
81}, 104431 (2010).

\bibitem{Cirac-Verstraete-2010} F. Verstraete and J. I. Cirac, Phys. Rev.
Lett. \textbf{104}, 190405 (2010).

\bibitem{Cirac-Verstraete-2004} F. Verstraete and J. I. Cirac,
cond-mat/0407066.

\bibitem{Sierra-1998} G. Sierra and M. A. Mart\'{\i}n-Delgado,
cond-mat/9811170.

\bibitem{Nishino-2001} T. Nishino, Y. Hieida, K. Okunishi, N. Maeshima, Y.
Akutsu, and A. Gendiar, Prog. Theor. Phys. \textbf{105}, 409 (2001).

\bibitem{Affleck-1987} I. Affleck, T. Kennedy, E. H. Lieb and H. Tasaki,
Phys. Rev. Lett. \textbf{59}, 799 (1987).

\bibitem{Haldane-1983} F. D. M. Haldane, Phys. Lett. \textbf{93}A, 464
(1983); Phys. Rev. Lett. \textbf{50}, 1153 (1983).

\bibitem{Arovas-1988} D. P. Arovas, A. Auerbach, and F. D. M. Haldane, Phys.
Rev. Lett. \textbf{60}, 531 (1988).

\bibitem{Sanz-2009} M. Sanz, M. M. Wolf, D. P\'{e}rez-Garc\'{\i}a, and J. I.
Cirac, Phys. Rev. A \textbf{79}, 042308 (2009).

\bibitem{Norbert-2008} N. Schuch, J. I. Cirac, and F. Verstraete, Phys. Rev.
Lett. \textbf{100}, 250501 (2008).

\bibitem{Bravyi-2006} S. Bravyi, quant-ph/0602108.

\bibitem{Shor-2010} R. Movassagh, E. Farhi, J. Goldstone, D. Nagaj, T. J.
Osborne, and P. W. Shor, Phys. Rev. A \textbf{82}, 012318 (2010).

\bibitem{Beaudrap-2010} N. de Beaudrap, M. Ohliger, T. J. Osborne, and J.
Eisert, Phys. Rev. Lett. \textbf{105}, 060504 (2010).

\bibitem{Laumann-2010} C. R. Laumann, R. Moessner, A. Scardicchio, and S. L.
Sondhi, Quantum Inf. Comput. \textbf{10}, 0001 (2010); C. R. Laumann, A. M. L%
\"{a}uchli, R. Moessner, A. Scardicchio, and S. L. Sondhi, Phys. Rev. A
\textbf{81}, 062345 (2010).

\bibitem{Kennedy-1992} T. Kennedy, J. Phys. A \textbf{25}, 2809 (1992).

\bibitem{Dukelsky-1998} J. Dukelsky, M. A. Mart\'{\i}n-Delgado, T. Nishino,
and G. Sierra, Europhys. Lett. \textbf{43}, 457 (1998).

\bibitem{McCulloch-2002} I. McCulloch and M. Gulacsi, Europhys. Lett.
\textbf{57}, 852 (2002).

\bibitem{Vidal-2009} S. Singh, H.-Q. Zhou, and G. Vidal, New J. Phys.
\textbf{12}, 033029 (2010); S. Singh, R. N. C. Pfeifer, and G. Vidal,
arXiv:0907.2994.

\bibitem{Brink-Satchler} D. M. Brink and G. R. Satchler, \textit{Angular
Momentum} (Clarendon Press, Oxford, 1968).

\bibitem{Tu-2008} H.-H. Tu, G.-M. Zhang, and T. Xiang, Phys. Rev. B \textbf{%
78}, 094404 (2008); J. Phys. A 41, 415201 (2008).

\bibitem{Tu-2009} H.-H. Tu, G.-M. Zhang, T. Xiang, Z.-X. Liu, and T.-K. Ng,
Phys. Rev. B \textbf{80}, 014401 (2009).

\bibitem{Scalapino-1998} D. Scalapino, S.-C. Zhang, and W. Hanke, Phys. Rev.
B \textbf{58}, 443 (1998).

\bibitem{Method} In practice, one can rule out some models to simplify the
calculation. For example, the element $2S$ is always included in $\mathcal{K}
$, otherwise the model has (at least) a fully polarized ferromagnetic ground
state. We also use the sets $\mathcal{K}^{\prime }$ for AKLT models and SO($%
2S+1$) models. Since these known models have unique ground states, there is
no need to check the sets $\mathcal{K}$ satisfying $\mathcal{K}\subset
\mathcal{K}^{\prime }$ or $\mathcal{K}\supset \mathcal{K}^{\prime }$.

\bibitem{Semi-integer} For semi-integer spin chains, it is not possible to
have a translationally invariant MPS as a unique ground state. According to
the Lieb-Schultz-Mattis theorem, SU(2) symmetric spin chains with
semi-integer spins are critical if the ground state is unique. On the
contrary, the fixed point MPS should have an energy gap and exponentially
decaying correlations. However, the possibility of the MPS solution with
breaking translational symmetry can not be ruled out. Although these models
are under the scope of exact renormalization method, we do not obtain any
such solution for $S\leq 7/2$.

\bibitem{Sanz-2010} M. Sanz, D. P\'{e}rez-Garc\'{\i}a, M. M. Wolf, and J. I.
Cirac, arXiv:0909.5347, to appear in IEEE Transactions on Information Theory.

\bibitem{Zheng-Zang-2010} D. Zheng, G.-M. Zhang, T. Xiang, and D.-H. Lee,
arXiv:1002.0171; J. Zang, H.-C. Jiang, Z.-Y. Weng, and S.-C. Zhang, Phys.
Rev. B \textbf{81}, 224430 (2010).

\bibitem{Rachel-2007} M. Greiter and S. Rachel, Phys. Rev. B \textbf{75},
184441 (2007); D. Schuricht and S. Rachel, Phys. Rev. B \textbf{78}, 014430
(2008); D. P. Arovas, K. Hasebe, X.-L. Qi, and S.-C. Zhang, Phys. Rev. B
\textbf{79}, 224404 (2009). K. Motegi, Phys. Lett. A \textbf{374}, 3112
(2010).
\end{thebibliography}
\end{document}